\documentclass[12pt,preprint]{aastex}

\shorttitle{A CLOUDY/XSPEC Interface}
\shortauthors{Porter et al.}

\begin{document}
\title{A CLOUDY/XSPEC Interface}


\author{R. L. Porter\altaffilmark{1}, G. J. Ferland\altaffilmark{1}, S. B. Kraemer\altaffilmark{2}, B. K. Armentrout\altaffilmark{2}, K. A. Arnaud\altaffilmark{3}, \& T. J. Turner\altaffilmark{4}}

\altaffiltext{1}{Dept. of Physics and Astronomy, University of Kentucky, Lexington, KY, 40506}
\altaffiltext{2}{Dept. of Physics, Catholic University of America, Washington, DC, 20064}
\altaffiltext{3}{Laboratory for X-ray Astrophysics, Code 662, NASA's Goddard Space Flight Center, Greenbelt, MD, 20771}
\altaffiltext{4}{Joint Center for Astrophysics, Physics Department, University of Maryland, Baltimore County, 1000 Hilltop Circle, Baltimore, MD 21250}



\begin{abstract}
We discuss new functionality of the spectral simulation code CLOUDY which allows the user to calculate grids with one or more initial parameters varied and formats the predicted spectra in the standard FITS format.  These files can then be imported into the x-ray spectral analysis software XSPEC and used as theoretical models for observations.  We present and verify a test case.  Finally, we consider a few observations and discuss our results.    
\end{abstract}

\keywords{methods: data analysis --- methods: statistical ---  X-rays: general}

\section{Introduction}
X-ray spectrometers typically record photon counts per energy bin.  The photon count for each bin is equal to the integral of the incident spectrum times an instrument response function, which is a function of both photon energy and detector bin.  In general, this integral cannot be reliably inverted to recover the incident spectrum, in part because inversion techniques tend to be unstable to small changes in the photon count.  The solution then is to compare the photon counts recorded by the spectrometer with best-fit theoretical photon counts, calculated by integrating the product of the known instrument response function with theoretical spectra.

XSPEC (last described by Arnaud 1996) is an open-source, X-ray spectral-fitting program, first developed in 1983 and still in development today.  The current \textit{de facto} standard for X-ray spectral analyis, XSPEC calculates theoretical photon counts from theoretical spectra to find the best fit between the observed and theoretical photon counts.  In order to do this, XSPEC must have not only data files containing the observed photon counts, background spectrum, and instrumental response (all of which can be readily obtained), but also theoretical models of any and all spectra that the user believes will accurately represent the actual source spectrum (as a whole or in part).  XSPEC comes bundled with many such theoretical models, but can also import external models via two methods: an external subroutine or a table file.  The table file option can be used if calculating an individual model is too CPU-intensive. The commands which control the importing of table models are as follows: \texttt{atable} for additive tabular models (intended to represent sources of emission), \texttt{mtable} for multiplicative tabular models (for absorption, filtering, and extinction), and \texttt{etable} for exponential tabular models (for any exponential effects).  The purpose of this work is to greatly expand the pool of theoretical models by creating functionality allowing the existing spectral simulation code CLOUDY to produce predicted spectra in a format which the user can import into XSPEC. CLOUDY predictions have been used in XSPEC before (recent examples include Kraemer et al. 2005 and Turner et al. 2005). However, the new functionality discussed here greatly simplifies the process.  We consider only additive and multiplicative models.

\section{The Spectral Simulation Code CLOUDY}

CLOUDY (last described by Ferland et al 1998) is an open-source, one-dimensional, spectral simulation code that has been in continuous development since 1978.  Researchers reference use of CLOUDY in over 100 publications a year.  The user gives commands specifying physical conditions such as intensity and shape of incident radiation, spatial and chemical composition, and pressure, density, and temperature laws, as well as commands controlling aspects of the microphysics, developmental features, or formatting of output.  CLOUDY then calculates a self-consistent solution and reports a predicted spectrum.   

CLOUDY has long had the ability to vary one of more initial parameters to try to find an optimal set of parameters to fit a specified emission-line spectrum, line flux or luminosity, and/or a set of column densities.  The \texttt{optimize} command and its keywords tell the code to vary one or more of the initial parameters to try to find an optimal set of parameters to fit a specified emission-line spectrum, line flux or luminosity, and/or a set of column densities.  It uses any of several minimization methods to obtain a best fit to a set of observed quantities.  The desired emission-line spectrum, line flux or luminosity, and/or column densities, are specified by a series of optimize commands.  A keyword \texttt{vary} can appear on several of the commands used to specify initial conditions to indicate which parameters are to be varied.  The commands with this option are listed in CLOUDY's documentation Hazy\footnote[1]{part 1, available at http://www.nublado.org/cloudy\underline{ }gold/hazy.htm}.  The optimization architecture lends itself well to computing predetermined grids of models; the code simply varies parameters in regular intervals (determined by the user) instead of in a feedback loop (dependent upon the results of all previous models).  
More options are planned for the future, including the ability to vary parameters in a second (or even third) spatial dimension.  The functionality discussed here will be supported in CLOUDY versions 6.04 and later.   

The commands used to control the XSPEC interface functionality are as follows:
\begin{itemize}
	\item \texttt{vary} - At least one command must have this option.  The value of the parameter specified on this line is ignored.
	\item \texttt{grid, range X to X} - Each command with the \texttt{vary} option must be followed by this command, which specifies the range of the parameter.  Both numbers X are required.    
	\item \texttt{grid, steps X} - This command must appear in order to actually turn on the grid functionality.  The other commands will be ignored if this command is not present.  The number X is optional and specifies the number of grid points for each variable.
	\item \texttt{punch xspec mtable "filename"} - Optional.  Produces a multiplicative table.
	\item \texttt{punch xspec atable "filename"} - Optional.  Produces an additive table.  This command has several options, as follows:
\begin{itemize}
	\item \texttt{[attenuated/reflected] incident continuum}
	\item \texttt{[reflected] diffuse continuum}
	\item \texttt{[reflected] lines}
	\item \texttt{[reflected] spectrum}
\end{itemize}
The transmitted portion will be selected or ``punched'' 
if neither \texttt{attenuated} nor \texttt{reflected} is included.  By ``reflected" we mean components escaping into the $2\pi$~sr subtended by the illuminated face toward the continuum source.  If no options are specified for the \texttt{atable} command, the transmitted spectra will be punched.  Note that the command can be specified multiple times, allowing the user to individually punch any or all of the separate components of the predicted spectra.  A redshift parameter is automatically added to all additive tables, but can be disabled easily in XSPEC.  See the discussion of energy binning below for limitations of including a redshift parameter. 
\end{itemize}
The files produced by the punch commands are 
in the FITS (Flexible Image Transport System) format.
FITS is a standard format used in astronomy, and endorsed by both NASA and the International Astronomical Union.
The most current definition of the FITS format is by Hanisch et al. (2001).
For more information on this format, visit the FITS Support Office website\footnote[2]{http://fits.gsfc.nasa.gov/fits\underline{ }home.html}.
A number of FITS image (and data) viewers and format converters are available\footnote[3]{http://fits.gsfc.nasa.gov/fits\underline{ }viewer.html}.
(The primary header of any FITS file produced by CLOUDY will list all the commands issued to CLOUDY in \texttt{COMMENT} tags.)
The table models used by XSPEC are a subset of the FITS format and are separately defined\footnote[4]{http://heasarc.gsfc.nasa.gov/docs/heasarc/ofwg/docs/general/ogip\underline{ }92\underline{ }009/ogip\underline{ }92\underline{ }009.html}.

The default energy resolution, $\Delta E/E$, used by CLOUDY is shown in Table \ref{table:resolution}.
At $1$~keV, the default resolution of 0.005 corresponds to a resolving power of 200, which will be better than the maximum resolving power achieved by the EPIC-MOS and EPIC-PN cameras on XMM-Newton, but comparable to or less than the maximum resolving power achieved with the Reflection Grating Spectrometer.
The resolving powers achieved by the Chandra X-Ray Observatory can be better still, especially with the Low Energy Transmission Grating Spectrometer and the High Energy Transmission Grating Spectrometer.
XSPEC will automatically rebin any table model to exactly match the bins of the specific instrument used in the observation, \textit{before} convolving the model with the instrument response function.  This rebinning introduces an uncertainty that can be made arbitrarily small by improving CLOUDY's resolution.  
The command \texttt{set continuum resolution XX} allows the user to modify the default resolution of CLOUDY by a constant factor (e.g., \texttt{set continuum resolution 0.1} will make the resolution ten times finer, or the resolving power ten times greater).
The user also has the option of defining a different continuum mesh, as described in Chapter 15 of Hazy 1.  This is recommended for analyzing any spectrum which includes $8.16$~keV ($600$~Ryd), as the default resolution becomes 6 times more coarse for energies just above that energy.  Note that a finer continuum mesh will be more CPU-intensive.

If a redshift parameter is used in XSPEC, it is important that the energy resolution of table models is significantly smaller than the redshift.  For example, at $1$~keV, the default resolution of 0.005 will only allow redshifts somewhat greater than 0.005 to be accurately treated.  Redshifts comparable to or less than the energy resolution would yield photon counts essentially indistinguishable from the photon count at zero redshift.  

\section{Examples}

As a test, we use CLOUDY to produce a simple blackbody spectrum, varying the temperature of the blackbody from $10^4$~K to $10^6$~K, with 21 logarithmically-spaced values (in $0.1$~dex steps).  We set the total luminosity to $10^{39}$~erg~s$^{ - 1}$ and the distance from the source to the illuminated face of the cloud to $10$~kiloparsecs.  According to the XSPEC manual, the normalization variable for an identical blackbody in XSPEC would be $K=L_{39}/D_{10}=1$, where $L_{39}$ is the luminosity in units $10^{39}$~erg~s$^{ - 1}$ and $D_{10}$ is the distance to the source in units of $10$~kpc.  The spectra produced by CLOUDY are then imported into XSPEC as additive tables.  

\begin{figure}
\includegraphics[angle=-90,width=6in,keepaspectratio=true]{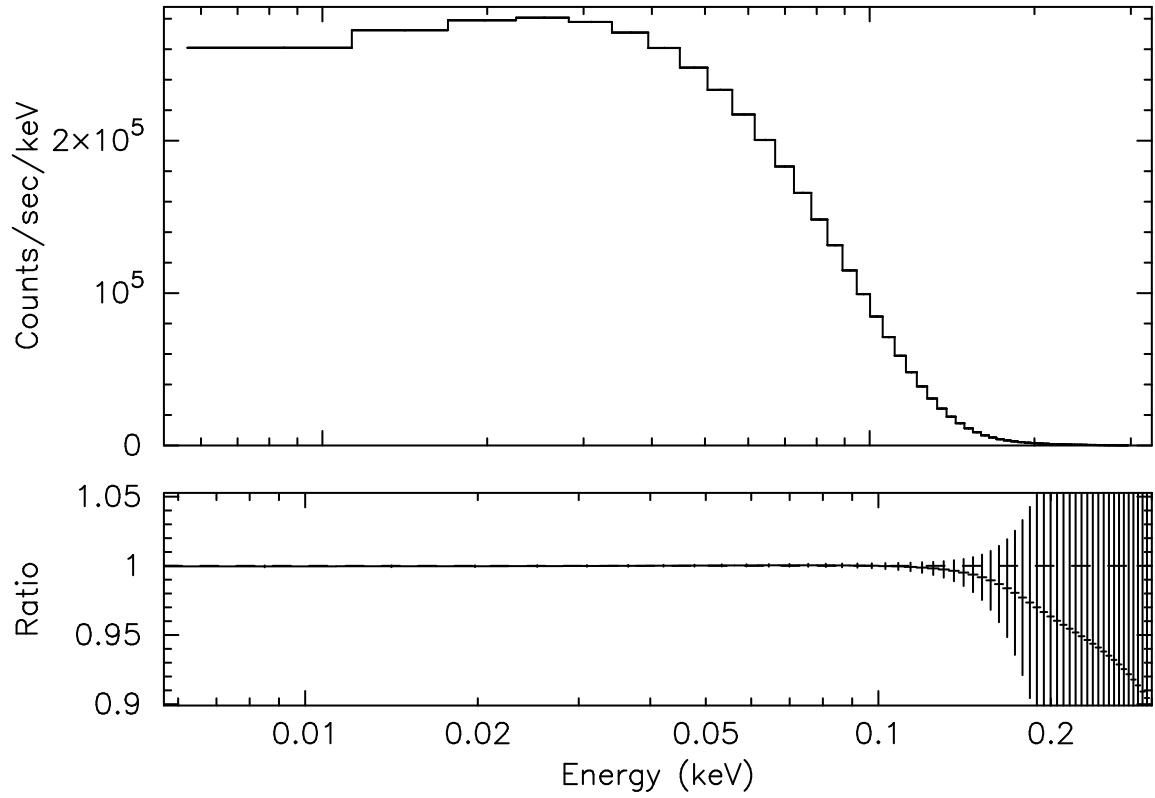}
\caption{A test of a blackbody model.  The data was derived from XSPEC's built-in \texttt{bbody} model using the \texttt{fakeit} command.  The folded model is the convolution of the chosen response matrix and the interpolated CLOUDY spectrum.}
\label{fig:blackbody}
\end{figure}

We then create a blackbody model in XSPEC with the command \texttt{model bbody} (setting $kT=0.02$~keV and $K=1.0$), and produce simulated photon counts of the model by convolving the spectrum with an instrument response matrix using the command \texttt{fakeit}.  (For the purposes of these test cases we answer ``no" to the question ``Use counting statistics in creating fake data?"  This allows us to isolate errors.)  For a perfect fit, we expect XSPEC to find $T=0.02$~keV$=232080$~K, or $log(T)=5.3656$, and normalization factor $norm=1$.  We impose initial guesses of $log(T)=5.0$ and norm$=2.0$, and XSPEC settles on $log(T)=5.35726$ and $norm=1.00460$.  If we decrease the spacing in our CLOUDY grid to $0.05$~dex steps, XSPEC finds $log(T)=5.36386$ and $norm=0.99297$.  If, on the other hand, we increase the spacing to $0.20$~dex steps, XSPEC finds $log(T)=5.33856$ and $norm=1.04767$.  Figure \ref{fig:blackbody} shows the fit (top-panel) and the ratio of data to folded model (bottom-panel) for the case with $0.05$~dex steps.  The decreasing ratio at higher energies is due to errors introduced by the interpolation and can be made arbitrarily small by further decreasing the grid spacing.      

\begin{figure}
\includegraphics[angle=-90,width=6in,keepaspectratio=true]{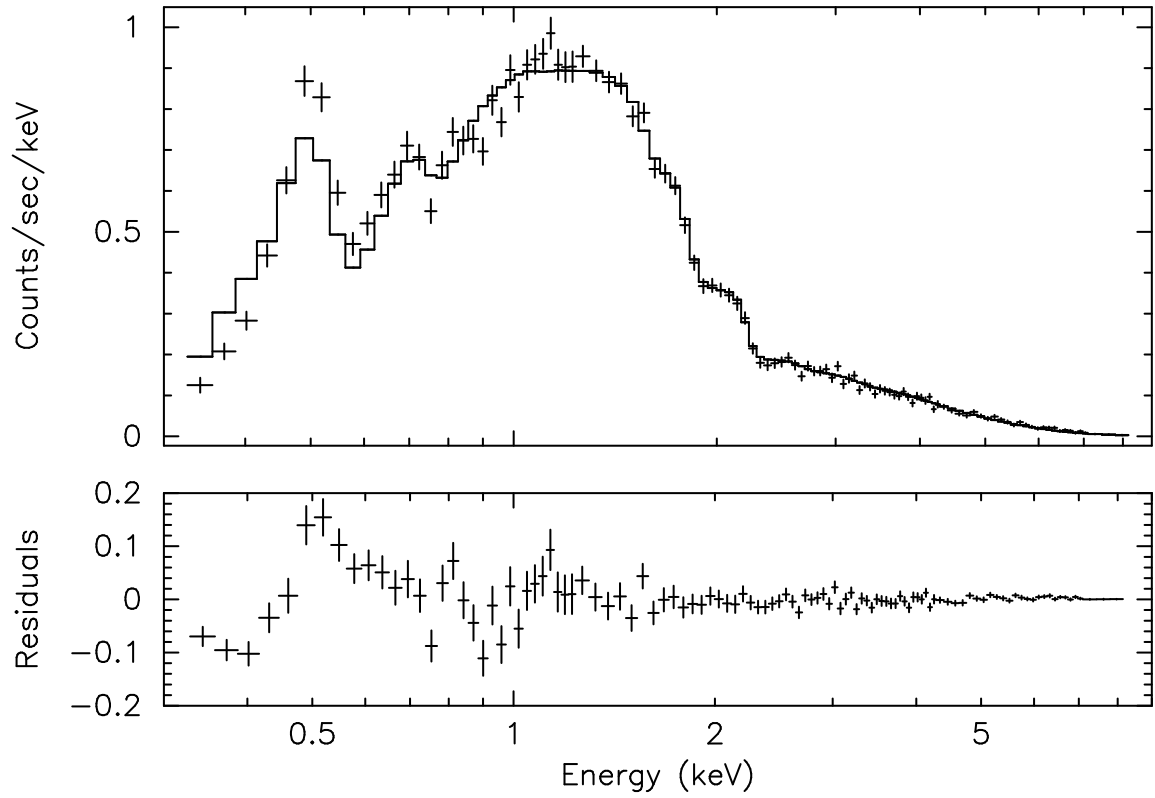}
\caption[]{An ASCA observation of MCG-6-30-15 (crosses) and the folded model (solid).  See text.}
\label{fig:mcg63015}
\end{figure}

To illustrate the use of a multiplicative table (which can not stand alone and must be multiplied by an additive table), we attempt to model the July 9, 1993, ASCA (SIS1) observation of MCG-6-30-15 (Fabian et al. 1994).  Following the procedure on page L61 of Fabian et al. (1994), we create a CLOUDY grid with a power law ``of photon index $\Gamma = 2$ and luminosity $10^{43}$~erg~s$^{-1}$, incident onto a shell at inner radius $10^{16}$~cm".  We then vary the ionization parameter and column density.  Fabian et al. then import the grid of transmitted spectra into XSPEC and find the best-fit column density, $N_\mathrm{w} = 10^{21.8}$~cm$^{-2}$, and ionization parameter, $\xi = 39$~erg~cm~s$^{-1}$ (corresponding to the dimensionless ionization parameter $U=2.38$).   Instead of using the transmitted spectra, we punch a multiplicative table, ``absorber.fit", which consists of the total $\exp(-\tau)$ from the illuminated face to the end of the calculation (determined here by the column density).  In XSPEC, we issue the command 
\begin{equation}
\texttt{model mtable$\left\{\mathrm{absorber.fit}\right\}$(powerlaw)}.
\end{equation}
With a total $\chi^2 = 259.7$ and a reduced $\chi^2 = 2.45$, XSPEC finds $\log N_\mathrm{w} = 21.82\pm0.03$~cm$^{-2}$, $\xi = 38.4^{+6.5}_{-5.6}$~erg~cm~s$^{-1}$, and $\Gamma = 2.11\pm0.03$.  The data and folded model are shown in figure \ref{fig:mcg63015}.  While our photon index is significantly greater than the $\Gamma = 2$ forced by Fabian et al., our other two parameters match their parameters to within the stated uncertainties.  Fabian et al. report a total $\chi^2 = 1081$, presumably including both July and August in the same fit, but do not report a reduced $\chi^2$.  Note that our model is simply for the demonstration of a multiplicative table and does not have the self-consistency of the Fabian et al. approach. 

\begin{figure}
\includegraphics[angle=0,width=6in,keepaspectratio=true]{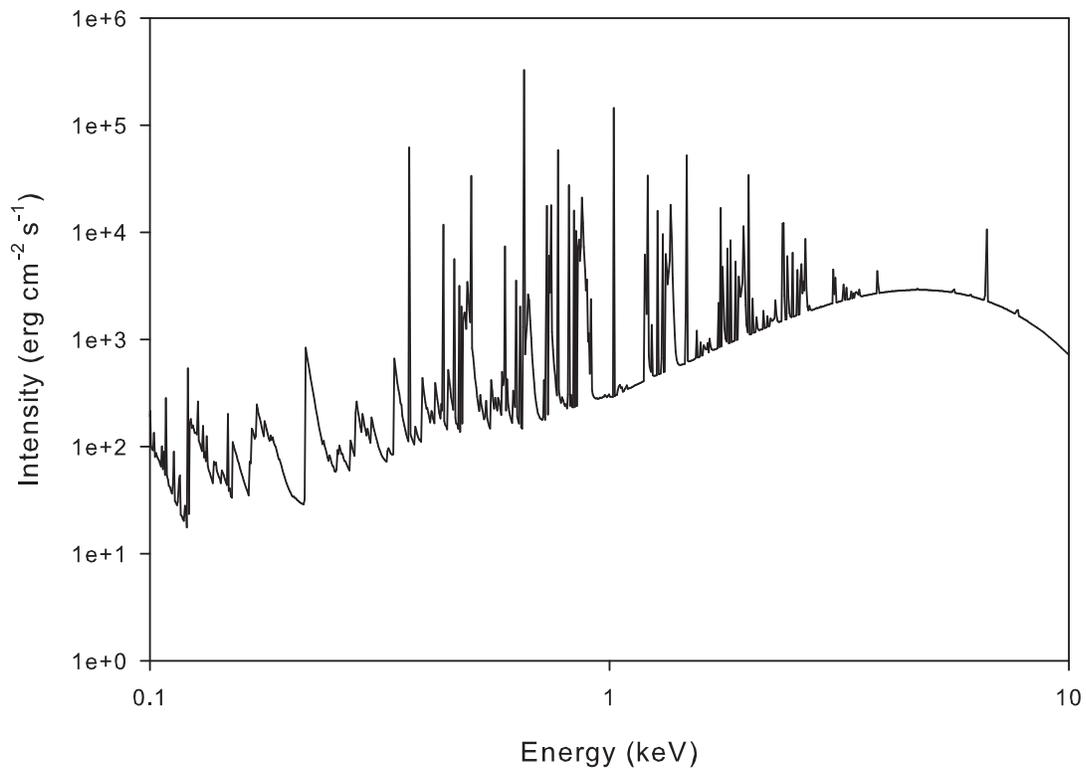}
\caption[]{A sample spectrum produced by CLOUDY, with the default energy resolution.  The radius parameter is 11.6, which means the illuminated face is $10^{11.6}$~cm from the center of the blackbody.  See text.}
\label{fig:cloudy_example}
\end{figure}

Finally, to illustrate an additive table with emission lines, we model the emission from the X-ray photoionized wind from binary Cygnus X-3.
Taking some cues from the appendix of Mitra (1996), but omitting the power law because we find it has little effect, we write the following CLOUDY commands:

\begin{verbatim}
blackbody 14,000,000
luminosity 38.48 range 0.1 keV to 10000 keV 
radius 11.5 vary
grid, range 11 to 12
wind 1000 km/sec
hden 13 
grid, steps 11
stop zone 1
punch last xspec atable spectrum "transpec.fit"
punch last xspec atable reflected spectrum "refspec.fit"
\end{verbatim}

A representative example of the produced reflected spectra is shown in figure \ref{fig:cloudy_example}.
In XSPEC, our model is built with the command
\begin{equation}
\texttt{model phabs(atable$\left\{\mathrm{transpec.fit}\right\}$+atable$\left\{\mathrm{refspec.fit}\right\}$)}.
\end{equation}
It is important to note that while the interpolation variables in our two tables are identical, and the two tables were produced by the same grid of models, XSPEC allows any variable to be a free parameter or tied to another parameter. In this case we force the radius (from the blackbody to the illuminated face) in our two tables to be tied but separate the normalization parameters.  The two normalization parameters then represent relative contributions from the transmitted and reflected spectra.

\begin{figure}
\includegraphics*[angle=-90,width=6in,keepaspectratio=true]{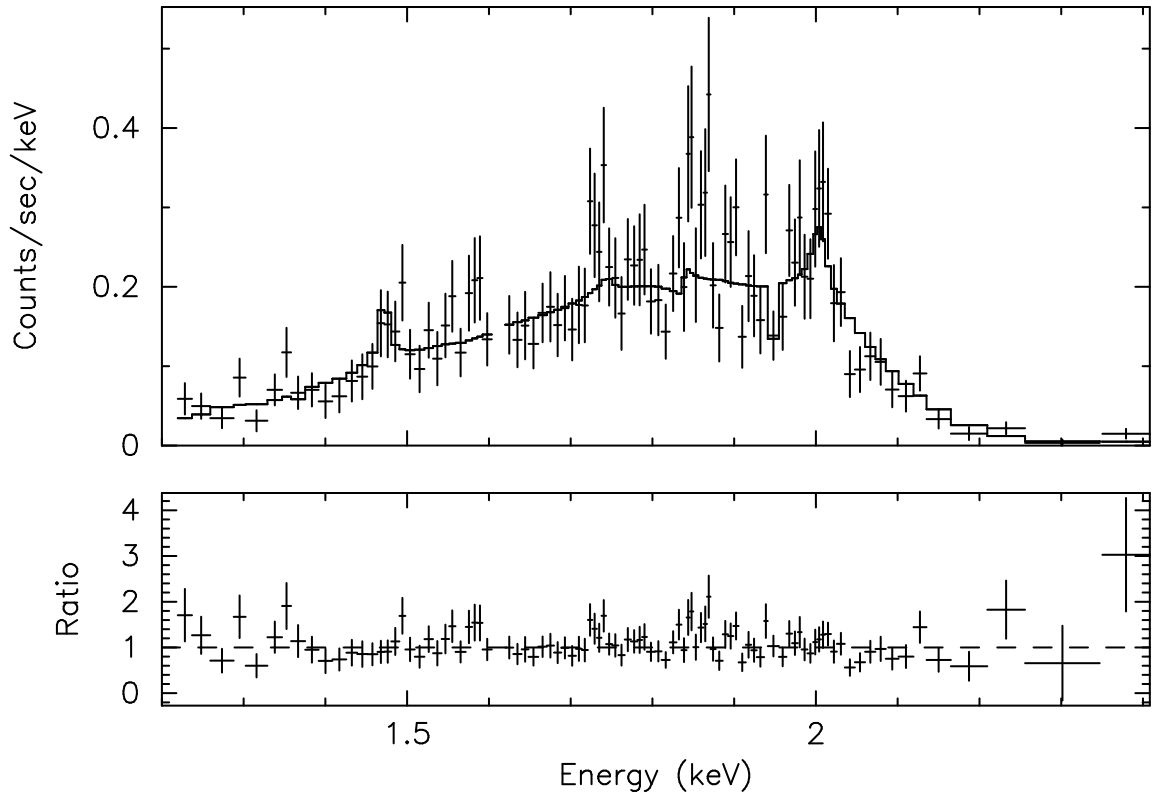}
\caption[]{Reduced data from an XMM-Newton observation of Cygnus X-3 (crosses) and our folded model (solid).}
\label{fig:cygnusx3_xmmn}
\end{figure}

We fit our model to the June 20, 2004, XMM-Newton observation of P.I. Martin Turner (observation ID 0165360101).
The observation has been filtered to include only time intervals with less than 7 counts per second and rebinned with a minimum of 20 counts per energy bin.  Both of these choices are somewhat arbitrary and based upon visual inspection of the data.
We consider only the first order spectrum from RGS1.  Our model fits the data with a reduced $\chi^2 = 1.09$.  The data and folded model are shown in figure \ref{fig:cygnusx3_xmmn}.  The normalization parameter for the reflected spectrum is orders of magnitude greater than the normalization parameter for the transmitted spectrum, indicating that the reflected spectrum provides a much better fit than does the transmitted spectrum.  As with the previous example, this is only for demonstration purposes.
A more detailed model would likely involve multiple zones and additional varied parameters.  


Both XSPEC and CLOUDY are freely available online, and both have accompanying documentation.  To download or find more details about XSPEC (or its parent software suite LHEASOFT), visit the NASA website at http://heasarc.gsfc.nasa.gov/docs/software.html.  To download or find more details about CLOUDY, visit its homepage at http://www.nublado.org.  

This research has made use of the Tartarus (Version 3.1) database, created by Paul O'Neill and Kirpal Nandra at Imperial College London, and Jane Turner at NASA/GSFC. Tartarus is supported by funding from PPARC, and NASA grants NAG5-7385 and NAG5-7067. We also acknowledge support from NASA grant NNG04GC56G.

\clearpage

\begin{deluxetable}{cr}
\tabletypesize{\scriptsize}
\tablecaption{The default resolution of CLOUDY. }
\tablewidth{0pt}
\tablehead
{
  \colhead{Energy Range}                               & 
  \colhead{Resolution ($\Delta E/E$)}												\\
}
\startdata
E $< 2.72\times10^{-7}~$keV	& 0.1 \\
$2.72\times10^{-7}~$keV~$< E < 8.16~$keV	&	0.005 \\
E $> 8.16~$keV	&	0.03 \\
\enddata
\label{table:resolution}
\end{deluxetable}

\clearpage

\clearpage

\clearpage

\clearpage

\end{document}